\documentclass{svmult12}

\usepackage{apjfonts}
\usepackage{helvet}         
\usepackage{courier}        
\usepackage{type1cm}        
\usepackage{graphicx}        
\usepackage{multicol}        
\usepackage[bottom]{footmisc}
\usepackage{url,hyperref}
\usepackage{natbib}
\usepackage{minaas-stuff}

\bibpunct[,]{(}{)}{;}{a}{}{,}
\newcommand{\kms}{km~s$^{-1}$}
\newcommand{\msun}{$M_\odot$}
\newcommand{\textdoi}
[1]{\href{http://dx.doi.org/#1}{DOI~\discretionary{}{}{}#1}}

\begin{document}

\title*{Type Iax Supernovae}
\author{Saurabh W.~Jha}
\institute{Saurabh W. Jha \at Department of Physics and Astronomy,
Rutgers, the State University of New Jersey, \\ 136 Frelinghuysen Road,
Piscataway, NJ 08854 USA, \\ \email{saurabh@physics.rutgers.edu}}
%
%
\maketitle

\vskip -1.5in
Author's version of a review chapter in the
\emph{Handbook of Supernovae}, \\ 
edited by Athem~W.~Alsabti and Paul~Murdin, 
Springer International Publishing, \\ 
\textdoi{10.1007/978-3-319-20794-0}

\vskip 0.7in
\abstract{Type Iax supernovae (SN~Iax), also called SN 2002cx-like
supernovae, are the largest class of ``peculiar'' white dwarf (thermonuclear)
supernovae, with over fifty members known. SN~Iax have lower ejecta velocity
and lower luminosities, and these parameters span a much wider range, than
normal type Ia supernovae (SN~Ia). SN~Iax are spectroscopically similar to
some SN~Ia near maximum light, but are unique among all supernovae in their
late-time spectra, which never become fully ``nebular''. SN~Iax overwhelmingly
occur in late-type host galaxies, implying a relatively young population. The
SN~Iax 2012Z is the only white dwarf supernova for which a pre-explosion
progenitor system has been detected. A variety of models have been proposed,
but one leading scenario has emerged: a type Iax supernova may be a
pure-deflagration explosion of a carbon-oxygen (or hybrid carbon-oxygen-neon)
white dwarf, triggered by helium accretion to the Chandrasekhar mass, that
does not necessarily fully disrupt the star.}

\vskip 0.7in
\section{Introduction}
\label{sec:intro}

\index{Type Iax supernova}
\index{Type Iax supernovae}
\index{SN Iax}
\index{SN 2002cx}
\index{SN 2002cx-like}
\index{02cx-like}
\index{peculiar supernovae}
Type Iax supernovae (SN~Iax) are a class of objects similar in some
observational properties to normal type-Ia supernovae (SN~Ia), but with clear
differences in their light-curve and spectroscopic evolution. SN~Iax are also
called ``02cx-like'' supernovae, based on the exemplar SN~2002cx, which was
described by \citet{Li03} as ``the most peculiar known'' SN~Ia. Later,
\citet{Jha06} presented other similar objects, making SN~2002cx ``the
prototype of a new subclass'' of SN~Ia. \citet{Foley13} coined the SN~Iax
classification, arguing these objects should be separated from SN~Ia as ``a
new class of stellar explosion.''

\index{SN 2005hk}
\index{SN 2008A}
\index{SN 2014ck}
Here I summarize the properties of SN~Iax, describing their observational
properties in Section \ref{sec:obs} and models in Section \ref{sec:models}. I
discuss analogues of SN~2002cx and well-studied examples like SN~2005hk
\citep{Chornock06,Phillips07,Stanishev07,Sahu08}, SN~2008A \citep{McCully14},
SN~2012Z \citep{Stritzinger15,Yamanaka15}, and SN~2014ck
\citep{2016MNRAS.459.1018T} as well as more extreme members of the class like
SN~2008ha \citep{Foley09,Valenti09} and SN~2010ae \citep{Stritzinger14}. There
may be some connection between SN~Iax and other classes of peculiar
white-dwarf supernovae \citep[like SN~2002es-like
objects;][]{Ganeshalingam12,White15,2016ApJ...832...86C} but I restrict my
focus to SN~Iax here; other peculiar objects are explored by
\citet{Taubenberger17}.

\section{Observations}
\label{sec:obs}

In this section I discuss the identification and classification of SN~Iax,
followed by their photometric and spectral properties from early to late time.
I also discuss the host environments of SN~Iax, their rates, and pre- and
post-explosion high-resolution imaging.

\subsection{Identification and Classification}
\label{sec:ident}

\begin{figure}
\centering \includegraphics[width=\textwidth]{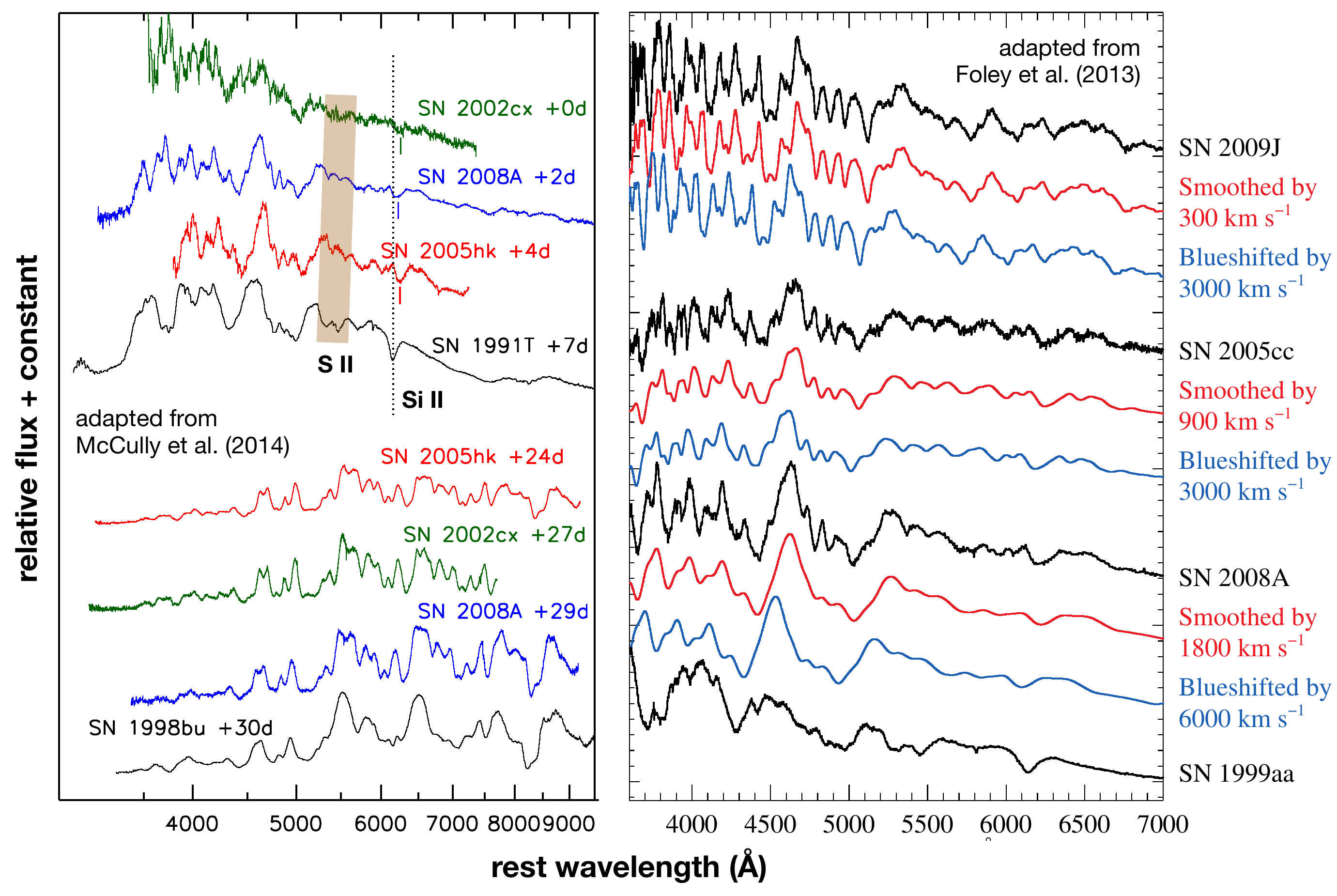}
\caption{Near-maximum light spectra of SN~Iax compared to normal and
91T/99aa-like SN~Ia. The left panel shows the similarity of SN~Iax to normal
SN~Ia (including the weak presence of \ion{S}{2} lines sometimes considered
hallmarks of thermonuclear SN), but also the lower expansion velocities (e.g.,
compare the locations of the \ion{Si}{2} lines). The right panel shows the
range of SN~Iax expansion velocities, from the lowest velocity SN~2009J
($v_{\rm exp} \approx 2200$ \kms) at the top, SN~2005cc ($v_{\rm exp} \approx
5000$ \kms) intermediate, and SN~2008A ($v_{\rm exp} \approx 6400$ \kms) at
the bottom, also compared to the normal SN~1999aa below. Note the similarity
in the spectra as the lower-velocity objects are smoothed and shifted to
resemble the higher-velocity objects. Figure adapted from \citet{McCully14}
and \citet{Foley13}.}
\label{fig:max-spectra}
\end{figure}

\index{supernova classification}
\index{SN Iax classification}
Supernovae are traditionally classified by their maximum light optical
spectra, and SN~Iax are no exception. These objects have spectra very similar
to some normal SN~Ia, particularly the ``hot'' SN~1991T-like or SN~1999aa-like
objects, with typically weak \ion{Si}{2} absorption and prominent \ion{Fe}{3}
lines \citep{Nugent95,Li01pecrate}. The key discriminant for SN~Iax is the
expansion velocity; unlike typical SN~Ia where the line velocity (usually
measured with \ion{Si}{2}) is $\sim$10,000 \kms, SN~Iax have much lower line
velocities anywhere from $6000-7000$ \kms\ (for objects like SN~2002cx,
2005hk, 2008A, 2012Z, and 2014dt) down to $2000$ \kms\ (for objects like
SN~2009J). Figure \ref{fig:max-spectra} shows maximum-light spectra of SN~Iax
with a range of expansion velocities, compared to normal SN~Ia. Because line
velocities are not always measured or reported when SN are classified and host
redshifts are sometimes uncertain, on occasion SN~Iax have been misclassified
as normal SN~Ia. Indeed, it is possible to identify unrecognized SN~Iax in
past observations, like SN~1991bj\footnote{One speculates how the history of
supernova cosmology would have changed if the diversity in SN~Ia were not only
typified by the extremes SN~1991T and SN~1991bg, but SN~1991bj as well, a
SN~Iax that would fall off the luminosity/light-curve decline rate
relationship!} \citep{Stanishev07}.  Table \ref{tab:list} gives a list of
53 objects that have been classified as SN~Iax.

\begin{table}
\caption{Partial list of type Iax supernovae. Host galaxies, redshifts,
and estimated peak absolute magnitudes are from the Open Supernova Catalog
\citep{OSC}. The references listed present the classification and/or
maximum-light data and are incomplete.}
\centering
\Scriptsize
\begin{tabular}{|c|c|c|c|c|}
\hline
$\,$ Supernova $\,$ & Host Galaxy & $z$ & $M_{\rm peak}$ & References
\\ 
\hline
SN 1991bj & IC 344 & $\,$ 0.018 $\,$ & $\, -15.3 \ $ &\citet{Stanishev07}\\
SN 1999ax & A140357+1551 & 0.023 & $\,-18.3\,$ & \citet{Foley13}\\ 
SN 2002bp & UGC 6332 & 0.020 & $-16.4$ & \citet{Silverman12} \\ 
SN 2002cx & CGCG 044$-$035 & 0.023 & $-18.8$ & \citet{Li03} \\ 
SN 2003gq & NGC 7407 & 0.021 & $-17.2$ & \citet{Jha06}\\ 
SN 2004cs & UGC 11001 & 0.014 & $-16.1$ & \citet{Foley13} \\ 
SN 2004gw & CGCG 283$-$003 & 0.017 & $-16.9$ & \citet{Foley09} \\ 
SN 2005P  & NGC 5468 & 0.009 & $-14.8$ & \citet{Jha06} \\ 
SN 2005cc & NGC 5383 & 0.007 & $-17.6$ & \citet{2005ATel..502....1A} \\ 
SN 2005hk & UGC 272 & 0.013 & $-18.5$ & \citet{Chornock06,Phillips07};\\
          & & & & \citet{Stanishev07,Sahu08}\\
SN 2006hn & UGC 6154 & 0.017 & $-18.9$ & \citet{Foley09}\\ 
SN 2007J  & UGC 1778 & 0.017 & $-17.2$ & \citet{2007CBET..926....1F}\\ 
SN 2007ie & SDSS J21736.67$+$003647.6 & 0.093 & $-18.2$ & \citet{2011AA...526A..28O}\\ 
SN 2007qd & SDSS J20932.72$-$005959.7 & 0.043 & $-16.2$ & \citet{McClelland10}\\ 
SN 2008A  & NGC 634 & 0.016 & $-19.0$ & \citet{McCully14}\\ 
SN 2008ae & IC 577 & 0.030 & $-18.8$ &$\,$\citet{2008CBET.1250....1B}$\,$\\
SN 2008ge & NGC 1527 & 0.003 & $-17.4$ & \citet{Foley10_ge}\\ 
SN 2008ha & UGC 12682 & 0.004 & $-14.0$ & \citet{Foley09,Valenti09} \\ 
SN 2009J  & IC 2160 & 0.015 & $-15.9$ & \citet{2009CBET.1665....1S}\\ 
SN 2009ho & UGC 1941 & 0.048 & $-18.2$ & \citet{2009CBET.1889....1S} \\
SN 2009ku & A032953$-$2805 & 0.079 & $-17.9$ & \citet{Narayan11}\\ 
PTF 09ego & SDSS J172625.23$+$625821.4 & 0.104 & $-18.6$ & \citet{White15}\\ 
PTF 09eiy & \nodata & 0.06 & \nodata & \citet{White15} \\
PTF 09eoi & SDSS J232412.96$+$124646.6 & 0.042 & $-16.7$ & \citet{White15} \\
SN 2010ae & ESO 162-G17 & 0.003 & $-14.0$ & \citet{Stritzinger14} \\ 
SN 2010el & NGC 1566 & 0.005 & $-13.0$ & \citet{2010CBET.2337....1B}\\ 
PTF 10xk  & \nodata & 0.066 & $-17.1$ & \citet{White15} \\
SN 2011ay & NGC 2315 & 0.021 & $-18.1$ & \citet{2015MNRAS.453.2103S} \\ 
SN 2011ce & NGC 6708 & 0.008 & $-17.1$ & \citet{2011CBET.2715....1M}\\ 
PTF 11hyh & SDSS J014550.57$+$143501.9 & 0.057 & $-18.7$ & \citet{White15} \\
SN 2012Z  & NGC 1309 & 0.007 & $-18.1$  & $\,$\citet{Stritzinger15,Yamanaka15} $\,$\\
PS1-12bwh & CGCG 205$-$021 & 0.023 & $-16.2$ &  \citet{Magee17}\\
LSQ12fhs  & \nodata & 0.033 & $-18.2$ & \citet{2012ATel.4476....1C}\\
SN 2013dh & NGC 5936 & 0.013 & $-17.3$ & \citet{2013ATel.5143....1J} \\
SN 2013en & UGC 11369 & 0.015 & $-17.9$ & \citet{2015MNRAS.452..838L}\\
SN 2013gr & ESO 114$-$G7 & 0.007 & $-15.4$ &\citet{2013ATel.5612....1H,2013CBET.3733....1H} \\
$\,$OGLE-2013-SN-130$\,$ & \nodata & 0.09 & $-18.0$ &\citet{2013ATel.5620....1B} \\
OGLE-2013-SN-147 & \nodata & 0.099 & $-19.3$ & \citet{2013ATel.5689....1L}\\
iPTF 13an & 2MASX J12141590$+$1532096 & 0.080 & \nodata & \citet{White15}\\
SN 2014ck & UGC 12182 & 0.005 & $-15.5$ & \citet{2016MNRAS.459.1018T} \\ 
SN 2014cr & NGC 6806 & 0.019 & $-17.0$ & \citet{2014ATel.6302....1C} \\ 
LSQ14dtt  &  \nodata & 0.05 & $-18.0$ & \citet{2014ATel.6398....1E} \\
SN 2014dt & NGC 4303 & 0.005 & $-18.6$ & \citet{Foley15,2016ApJ...816L..13F} \\
SN 2014ek & UGC 12850 & 0.023 & $-18.0$ & \citet{2014ATel.6611....1Z}\\ 
SN 2014ey & CGCG 048$-$099 & 0.032 & $-18.1$ & \citet{2017arXiv170704270L}\\
SN 2015H  & NGC 3464 & 0.012 & $-17.7$ & \citet{Magee16}\\
PS15aic   & $\,$2MASX J13304792$+$3806450$\,$ & 0.056 & $-17.9$ &\citet{2015ATel.7534....1P} \\
SN 2015ce & UGC 12156 & 0.017 & $-17.9$ & \citet{2017TNSCR.381....1B} \\ 
PS15csd   & \nodata & 0.044 & $-17.5$ & \citet{2015ATel.8264....1H} \\
SN 2016atw & \nodata & 0.065 & $-18.0$ & \citet{2016ATel.8810....1P} \\ 
OGLE16erd & \nodata & 0.035 & $-17.1$ & \citet{2016ATel.9660....1D} \\ 
SN 2016ilf & 2MASX J02351956$+$3511426 & 0.045 & $-17.6$ & \citet{2016ATel.9795....1Z}\\ 
iPTF 16fnm & UGC 00755 & 0.022 & $-15.0$ & \citet{2017arXiv170307449M} \\
\hline
\end{tabular}
\label{tab:list}
\end{table}

\subsection{Photometric properties}
\label{sec:photprops}

The optical light curves of SN~Iax show a general similarity to SN~Ia, though
with more diversity. SN~Iax typically have faster rises ($\sim$10 to 20 days)
in all bands, with pre-maximum light curves showing significant variety
\citep{Magee16,Magee17}. The $B$ and $V$-band decline rates are similar to
normal SN~Ia, though generally also on the faster side \citep{Stritzinger15},
and the optical color evolution in SN~Iax (e.g., in $B-V$) has a shape roughly
similar to normal SN~Ia \citep{Foley13}. However, SN~Iax have significantly
slower declines in redder bands, e.g., $\Delta m_{15}(R) \simeq$ 0.2 to 0.8
mag, compared to normal SN~Ia which have $\Delta m_{15}(R) \simeq$ 0.6 to 0.8
mag \citep{Magee16}. Faster rising SN~Iax are generally faster fading as well,
with some exceptions, like SN~2007qd \citep{McClelland10}. SN~Iax do not show
the prominent ``second-peak'' in the redder and near-infrared bands
\citep{2014ApJ...795..142G} that characterize normal SN~Ia. The second-maximum
is especially strong in slowly-declining SN~1999aa or 1991T-like SN~Ia; thus
the SN~Ia that are spectroscopically most similar to SN~Iax have quite a
different photometric behavior. Nonetheless, similar to SN~Ia, weeks after
maximum light SN~Iax show only modest color evolution. Late-time colors may
thus provide a useful diagnostic of host-galaxy reddening
\citep{Lira96,Foley13}, which is otherwise difficult to determine for SN~Iax.
The very late-time optical light curves of typical SN~Iax continue to show a
decline slower than SN~Ia until about 300 to 400 days past maximum light,
after which SN~Ia light curves also slow to similar decline rates as SN~Iax:
$0.01-0.02$ mag day$^{-1}$ \citep{McCully14}.

\begin{figure}
\centering \includegraphics[width=0.9\textwidth]{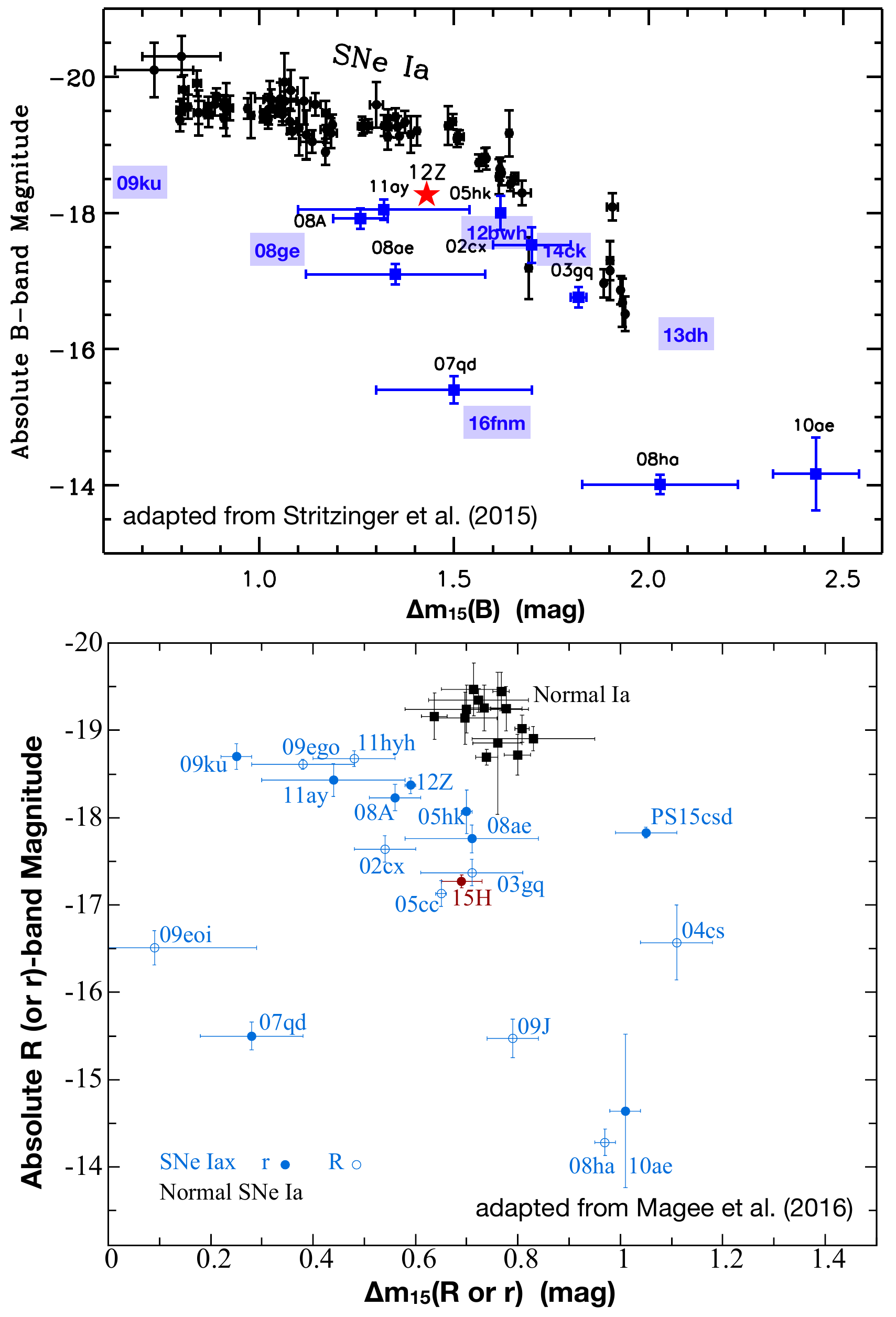}
\caption{Absolute magnitude vs. decline rate relation for SN~Iax 
(colored) compared to normal SN~Ia (black) showing the \citet{Phillips93}
relation, in $B$-band (above) and $R$ (or $r$)-band below. These plots are
adapted from \citet{Stritzinger15} and \citet{Magee16}.}
\label{fig:phillips}
\end{figure}

The peak optical luminosity of SN~Iax is lower than typical SN~Ia and it spans
a much wider range, from $M_V \simeq -19$ on the bright end to $M_V \simeq
-13$ for the faintest SN~Iax known. Compared to the light curve decline rate,
SN~Iax fall well below the \citet{Phillips93} relation, by anywhere between
0.5 and several magnitudes \citep{Foley13}. Certainly, as seen in Figure
\ref{fig:phillips}, SN~Iax do not show as tight a relation in this parameter
space as normal SN~Ia (even including the SN 1991T/1999aa and SN 1991bg
extremes of the normal SN~Ia distribution). \citet[ see their Figure
5]{Magee16} suggest a stronger correlation may exist between peak luminosity
and rise time, rather than decline rate.

Near-infrared light curves of SN~Iax are limited, with SN~2005hk still
providing the best data set \citep{Phillips07}. Continuing the trend with
wavelength in the optical, the $YJH$ light curves of SN~2005hk show a broad,
single peak that is delayed significantly ($\sim$10 to 15 days) relative to
the $B$ peak. The NIR contribution to the quasi-bolometric ``UVOIR'' flux
seems not too dissimilar to normal SN~Ia \citep{Stritzinger15}, and this
fraction seems roughly consistent even for the faintest SN~Iax like SN~2008ha
and SN~2010ae \citep{Stritzinger14}. A major surprise, however, is SN~2014dt,
which showed a significant near- and mid-infrared excess beginning $\sim$100
days past maximum and lasting for a few hundred days at least
\citep{2016ApJ...816L..13F}.

The near-UV photometric behavior of SN~Iax is also interesting, with objects
showing a faster evolution in near-UV minus optical color than SN~Ia, and
``crossing'' the typically parallel tracks made by normal SN~Ia in this space.
SN~Iax start bluer than normal SN~Ia in the UV before maximum light but
quickly redden (by $\sim$ 1.5 to 2 mag in Swift $uvw1 - b$) so that about ten
days after maximum they are redder than normal SN~Ia
\citep{2010ApJ...721.1627M}.

As with other thermonuclear supernovae, no SN~Iax has been definitively
detected in the radio \citep{2016ApJ...821..119C} or X-ray
\citep{2014ApJ...790...52M}.

\subsection{Spectroscopic properties}
\label{specprops}

Beyond the defining spectroscopic features used for classification of these
supernovae (\ion{Fe}{3} dominated spectra near maximum light and low
\ion{Si}{2} velocity; see Sec. \ref{sec:ident}), SN~Iax show quite
homogeneous spectral evolution, which generally matches the evolution of SN~Ia
over the period of a few months from maximum light, except with lower line
velocities \citep{Jha06}. Like SN~Ia, the early-time spectra show Fe-group and
intermediate mass elements (including Si, S, and Ca). This similarity extends
to the near-UV (with Fe-group line blanketing). In their near-infrared
maximum-light spectra, SN~Iax show remarkable similarity to SN~Ia, with
\ion{Fe}{2} and \ion{Si}{3} lines, except at lower typical velocities, and
most prominently, beautiful and unambiguous detections of
\ion{Co}{2} in the $H$ and $K$ bands for SN~2010ae, 2012Z, and 2014ck
\citep{Stritzinger14,Stritzinger15,2016MNRAS.459.1018T}.

\begin{figure}
\includegraphics[width=\textwidth]{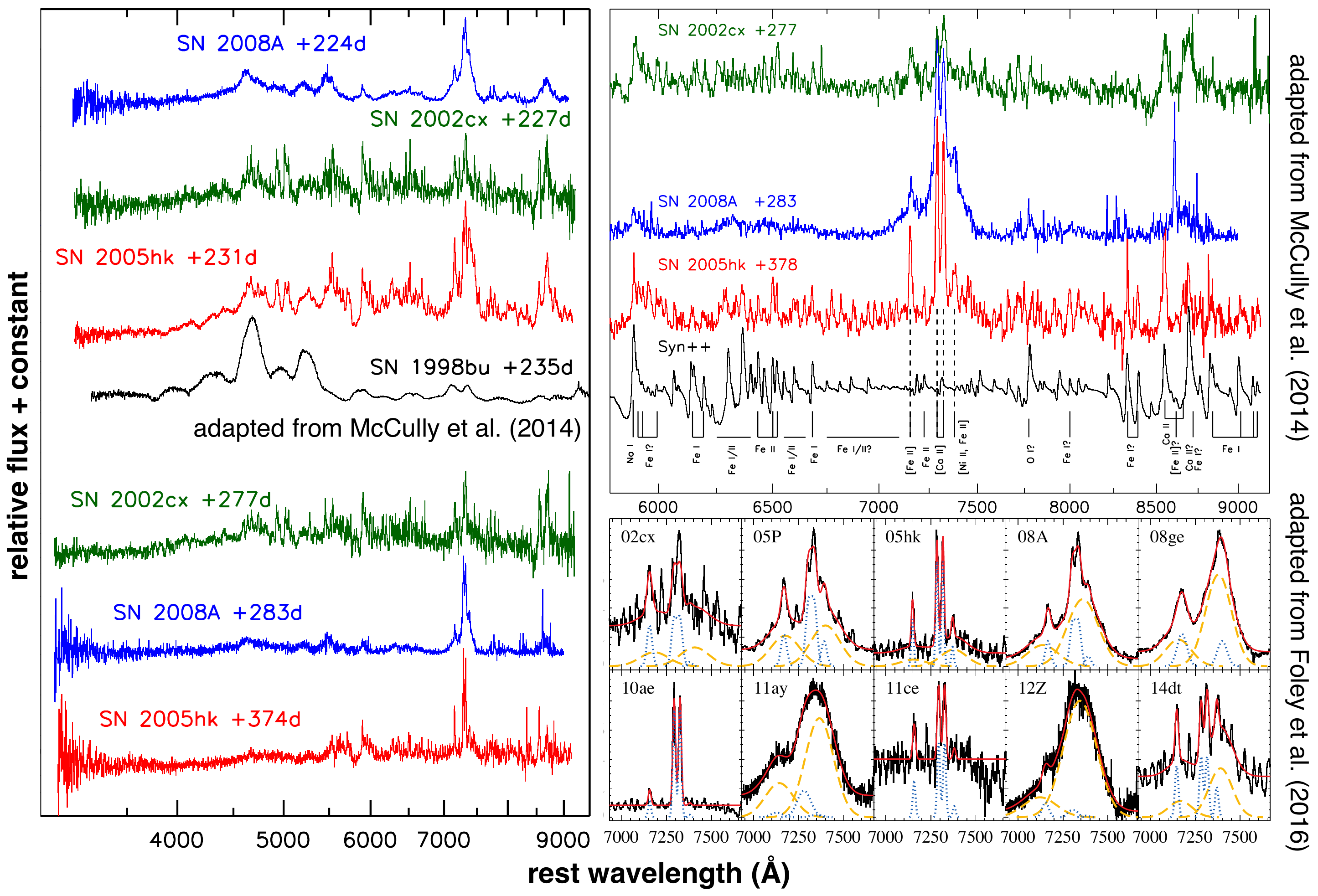}
\caption{Late-time spectra of SN~Iax (colored) compared to
  a normal SN~Ia (black; left), showing a clear divergence. Numerous lines
  that look like ``noise'' are actually permitted Fe transitions, as seen with
  the Syn++ \citep{Thomas11_syn} spectrum synthesis model (black) of SN~2005hk
  (upper right). SN~Iax show a diversity of line velocities and strengths in
  the [Fe II] $\lambda 7155$, [Ca II] $\lambda\lambda 7291,7324$, and [Ni II]
  $\lambda 7378$ lines at late times. Figure panels adapted from
  \citet{McCully14} and \citet{Foley16}.}
\label{fig:late}
\end{figure}

It is at late times that the spectra of SN~Iax radically diverge from SN~Ia,
and indeed, almost all other supernovae of any type
\citep{Jha06,Sahu08,Foley10,Foley16}. SN~Iax never truly enter a fully
``nebular'' phase in which broad forbidden lines dominate the optical spectrum
(Figure~\ref{fig:late}). Rather, in optical spectra taken more than a year
past maximum light, SN~Iax still show permitted lines of predominantly
\ion{Fe}{2}, often with low velocities $<$ 2000 \kms, plus \ion{Na}{1} D and
the \ion{Ca}{2} IR triplet. Forbidden lines of [\ion{Fe}{2}], [\ion{Ni}{2}],
and [\ion{Ca}{2}] are also usually present, and in some cases with narrow
widths down to $<$ 500 \kms\ \citep{McCully14,Stritzinger15}. The linewidths
and relative strengths of the forbidden and permitted lines seem to vary
significantly among different SN~Iax \citep{Yamanaka15,Foley16}. The late-time
linewidth variation, for example, approaches nearly an order of magnitude,
from a few hundred to $\sim$3000 \kms. Spectra of SN~Iax seem to show
relatively little evolution from 200 to past 400 days after maximum
\citep{Foley16}.

\index{SN 2008ha}
\index{SN 2010ae}
The faintest SN~Iax, like SN~2008ha and SN~2010ae, show similar spectra to
more luminous counterparts, perhaps with a more rapid spectral evolution to
lower velocities in the few weeks after maximum light
\citep{Foley09,Valenti09,Stritzinger14}. Around 250 days past maximum, the
spectra of SN~2002cx \citep[a ``bright'' SN~Iax;][]{Jha06} and SN~2010ae (one
of the faintest) are nearly identical \citep[see Figure 12
of][]{Stritzinger14}.

\citet{Chornock06} and \citet{Maund10} obtained spectropolarimetric
observations of SN~2005hk and report 0.2--0.4\% continuum polarization,
consistent with spectropolarimetry of normal SN~Ia. 

\index{helium in SN Iax}
Two objects, SN~2004cs and 2007J, have been classified as SN~Iax by
\citet{Foley13} and show clear evidence of \ion{He}{1} emission in their
post-maximum spectra, something never seen in normal SN~Ia. \citet{White15}
argue that these objects may be type-IIb supernovae instead, though
\citet{Foley16} counter that claim and call PTF 09ego and PTF 09eiy into
question as SN~Iax. In the end, there are a handful of objects for which the
classification may be ambiguous.

\subsection{Photometric and spectroscopic correlations}
\label{photspec}

SN~Iax span a wide range of peak luminosities and line velocities, and it is
natural to ask if these are related. \citet{McClelland10} suggested a positive
correlation between these two, with the lowest-velocity SN~Iax also being the
lowest luminosity. While such a correlation does seem to hold for the majority
of SN~Iax, there are clear counterexamples like SN~2009ku \citep{Narayan11},
which had a low velocity (similar to SN~2008ha or SN~2010ae), but relatively
high luminosity (like SN~2002cx or SN~2005hk). \citet{2016MNRAS.459.1018T}
show that SN~2014ck is similarly an outlier to the velocity/luminosity
correlation. Neither do the lowest-velocity SN~Iax necessarily have the
fastest optical decline rates: while SN~2008ha and SN~2010ae decline quickly,
SN~2014ck has an intermediate decline rate similar to other higher-velocity
SN~Iax, while SN~2009ku in fact has a slower decline rate than higher-velocity
SN~Iax.

\subsection{Environments and Rates}
\label{environments}

\begin{figure}
\includegraphics[width=\textwidth]{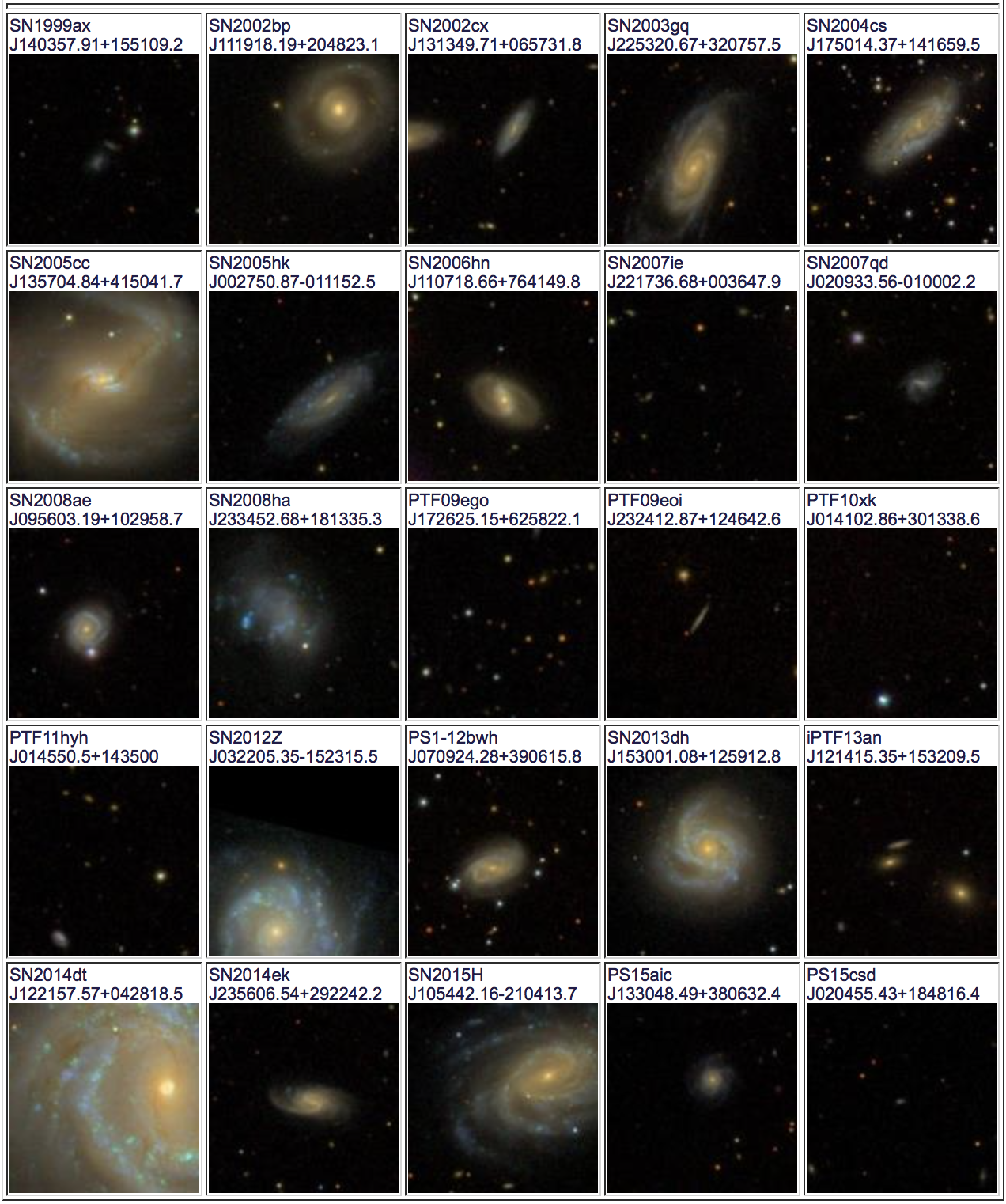}
\caption{Sloan Digital Sky Survey images centered at the locations of 25
SN~Iax in the SDSS footprint. Note the preponderance of late-type,
star-forming host galaxies. Each image is 100 arcsec on a side.}
\label{fig:gallery}
\end{figure}

\index{host galaxies}
\index{SN Iax environments}
SN~Iax have a dramatically different distribution of host galaxies than normal
SN~Ia. In all but a couple of cases \citep[like SN~2008ge;][]{Foley10_ge},
SN~Iax are found in star-forming, late-type host galaxies (Figure
\ref{fig:gallery}). The host galaxy distribution of SN~Iax is closest to those
of SN~IIP or SN 1991T/1999aa-like SN~Ia \citep{Foley09,Valenti09,Perets10}.
\citet{Lyman13,2017arXiv170704270L} confirm and amplify this result based on 
H$\alpha$ imaging and integral-field spectroscopy of
Iax locations and hosts: SN~Iax must arise from a relatively young population.
Qualitatively, based on SN~Iax with high-resolution Hubble Space Telescope
like SN~2008A, SN~2012Z, and SN~2014dt, it seems as if SN~Iax prefer the
``outskirts'' of their star-forming hosts, but this needs further
quantification, given the selection bias against finding these fainter SN on a
bright galaxy background. SN~Iax show no strong preference for high- or
low-metallicity galaxies \citep{Magee17}, though their explosion locations
are more metal-poor than normal SN~Ia \citep{2017arXiv170704270L}.

\index{SN 2013en}
The host reddening distribution of SN~Iax is uncertain because of the
difficulty in disentangling extinction from the intrinsic photometric
diversity in the class, but most known SN~Iax have low reddening, up to
$E(B-V) \simeq 0.5$ mag for SN~2013en \citep{2015MNRAS.452..838L}. Again,
selection biases work against finding heavily extinguished members of this
already intrinsically faint class of supernovae.

\index{SN Iax rate}
Because the luminosity function of SN~Iax extends down to quite faint
magnitudes, precisely estimating the rate of SN~Iax is challenging.
\citet{Foley13} calculate the Iax rate to be 31$^{+17}_{-13}$\% of the SN~Ia
rate in a volume-limited sample. Consistent with this,
\citet{2017arXiv170307449M} found one SN~Iax (iPTF16fnm; $M \simeq -15$ mag)
and 4 SN~Ia in a volume-limited survey. The overall SN~Iax rate is dominated
by the lower luminosity objects; the rate of brighter SN~Iax (comparable in
luminosity to SN~2002cx or SN~2005hk) is likely to be between 2 and 10\% of
the SN~Ia rate \citep{2011MNRAS.412.1441L,Foley13,2017ApJ...837..121G}. SN~Iax
are the most numerous ``peculiar'' cousins to normal SN~Ia.

\subsection{Progenitors and Remnants}

\begin{figure}
\includegraphics[width=\textwidth]{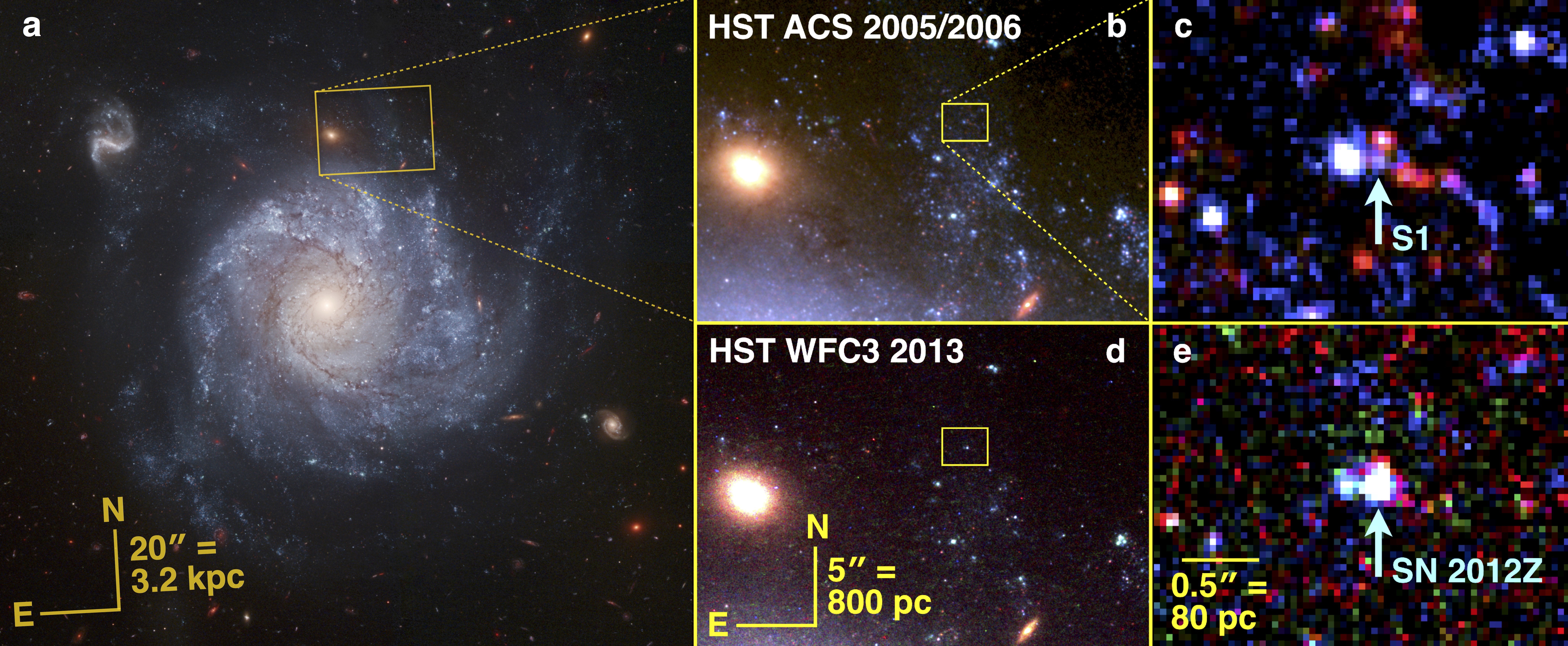}
\caption{Discovery of the only known progenitor system
for a white-dwarf (thermonuclear) supernova. The left panel shows the deep
Hubble Heritage image of NGC 1309 (made from observations to detect and
monitor Cepheids) taken in 2005 and 2006 with HST ACS. Upper panels (b) and
(c) zoom in on the region where the type Iax SN~2012Z was to explode,
revealing the luminous, blue progenitor system S1, believed to be a helium
star donor to an exploding white dwarf. Lower panels (d) and (e) show the
region after the SN explosion, allowing for a precise measurement of its
position with HST/WFC3, coincident with the progenitor. This figure is adapted
from \citet{McCully14_12Z}.}
\label{fig:12Z}
\end{figure}

\index{SN 2012Z}
A major breakthrough in understanding SN~Iax came with the discovery of the
progenitor system of SN~2012Z \citep{McCully14_12Z}. Nature was kind: SN~2012Z
exploded in the nearby galaxy NGC~1309, which was also the host of the normal
type-Ia SN~2002fk, a calibrator for the SN distance scale to measure $H_0$
\citep{Riess11}. As such, extremely deep, multi-epoch HST imaging of NGC~1309
(to observe Cepheids) covered the location of SN~2012Z before its explosion
(Figure \ref{fig:12Z}). This deep, high-resolution pre-explosion imaging
revealed a source coincident with SN~2012Z, the first time a progenitor system
has been discovered for a thermonuclear supernova. The detected source is
luminous and blue ($M_V \simeq -5.3$ mag; $B-V \simeq -0.1$ mag), and
\citet{McCully14_12Z} argue that it is a helium-star companion (donor) to an
exploding white dwarf. Further images taken after the supernova faded reveal
the source has not disappeared, consistent with the companion scenario
(McCully et al., in preparation). This discovery marks a critical contrast for
SN~Iax compared to normal SN~Ia: no such progenitor system has ever been seen
for a SN~Ia! Note, however, there are only two normal SN~Ia, SN~2011fe
\citep{Li11} and SN~2014J \citep{Kelly14}, with pre-explosion limits that are
as deep as the data for SN~2012Z.

\citet{Foley14} detect a luminous ($M_I \simeq -5.4$ mag), red ($R-I \simeq
1.6 \pm 0.6$ mag) source consistent with the location of SN~2008ha in HST
imaging taken about 4 years after the supernova explosion. At these epochs the
SN ejecta flux should have faded well below this level, so \citet{Foley14}
suggest they may be observing a companion star to the supernova, or else a
luminous ``remnant'' of the explosion.

\section{Models}
\label{sec:models}

In this section I discuss potential models for SN~Iax, starting from
general considerations and observational constraints, and then moving to
specific scenarios that have been proposed in the literature. 

\subsection{SN~Iax are likely thermonuclear, white dwarf supernovae}

SN~Iax show undeniable spectroscopic similarities to normal SN~Ia,
particularly near and in the few months after maximum light, with lower
velocities being the primary distinguishing factor. Given that spectroscopic
observations probe different layers of supernova ejecta over
time\footnote{``Spectrum is truth.'' --R.~P.~Kirshner}, a natural starting
point would be to suggest that SN~Iax and SN~Ia share commonalities in their
progenitors and explosions.

Conversely, the primarily star-forming environments of SN~Iax may point to a
core-collapse, massive-star supernova origin. Indeed \citet{Valenti09} argue
that SN~2008ha has similarities to some faint core-collapse SN, and suggest a
``fallback'' massive-star supernova \citep{Moriya10} or an electron-capture
supernovae \citep{2009ApJ...705L.138P} could explain SN~2008ha, though not
without some difficulties \citep{2013MNRAS.436..774E}. This core-collapse
model does not seem to be able to account for higher luminosity SN~Iax, and so
would require objects like SN~2008ha, 2010ae, and 2010el to be distinct from
other SN~Iax.

The preponderance of evidence suggests that SN~Iax are thermonuclear
explosions of white dwarfs. Spectra near maximum light show carbon,
intermediate mass elements like sulfur and silicon (weakly), and strong
features of iron group elements. The \ion{Co}{2} infrared lines
\citep{Stritzinger15,2016MNRAS.459.1018T} clearly point to a fraternity with
normal SN~Ia. Iron lines are seen at a wide range of velocities, implying
efficient mixing of fusion products rather than a highly layered structure.
The lack of star formation or luminous massive stars in pre-explosion imaging
of SN~2008ge \citep{Foley10_ge} and SN~2014dt \citep{Foley15}, and the
non-disappearance of the progenitor system flux in SN~2012Z
\citep{McCully14_12Z} argue against the explosion of massive luminous stars.
The peak luminosities of SN~Iax compared to their late-time photometry and
modeling suggest a $^{56}$Ni $\rightarrow\;^{56}$Co $\rightarrow\;^{56}$Fe
radioactively powered light curve \citep{McCully14}.

Moreover, even fainter, lower velocity objects like SN~2008ha and 2010ae
seem to connect to brighter SN~Iax. \citet{Foley10} show an early
spectrum of SN~2008ha that features sulfur lines similar to those seen in
normal SN~Ia. \citet{Stritzinger14} show that SN~2010ae has strong
\ion{Co}{2} lines like other SN~Iax and normal SN~Ia, and its late-time
spectrum is very similar to the SN~Iax prototype, SN~2002cx.

\subsection{General observational constraints}

The environments of SN~Iax do indeed suggest they come from a young
population, but this does not require a core-collapse origin. SN~1991T-like
SN~Ia have a quite similar host galaxy preference \citep{Foley09,Perets10} to
SN~Iax, and those SN~Ia are still nearly-universally construed as white dwarf
explosions. In fact, the problem can be turned around: the requirement for a
young population favors certain binary systems that can produce and explode
white dwarfs quickly. HST observations of nearby stars in the field of
SN~2012Z yield ages of 10--50 Myr \citep{McCully14_12Z}, while for SN~2008ha
the nearby population is $\lesssim$ 100 Myr \citep{Foley14}. Though young,
these are still significantly older than the expected lifetimes of, for
example, Wolf-Rayet stars that might yield hydrogen-poor core-collapse
supernovae \citep{Groh13}.

Short evolutionary times in a binary system suggest that SN~Iax arise from
more massive white dwarfs. If the explosions are occurring at the
Chandrasekhar mass ($M_{\rm Ch}$), as supported by other evidence (see below),
then the quickest binary channel for a carbon/oxygen white dwarf (C/O WD) is
to accrete helium from a He star companion
\citep{1999ApJ...519..314H,2014LRR....17....3P}. The stable mass transfer rate
can be high for helium accretion, and \citet{Claeys14} show that this channel
dominates the thermonuclear SN rate between 40 Myr (with no younger systems)
and 200 Myr (above which double-degenerate and hydrogen-accreting single
degenerate systems dominate). This has been seen in several binary population
synthesis studies; these generally find no problem for the He star + C/O WD
channel to produce the required fraction of SN~Iax relative to normal SN~Ia,
but the total rates may not quite reach the observed values \citep{Ruiter09,
2011MNRAS.417..408R,2009ApJ...701.1540W,2009MNRAS.395..847W,
2010ApJ...710.1310M,2014MNRAS.445.3239P,2015A&A...574A..12L}.

Of course the WD+He star channel is in good accord with observations
of SN~2012Z, for which the putative companion is consistent with a
helium star. This scenario may also explain the helium observed in
SN~2004cs and SN~2007J. In fact, \citet{Foley13} predicted this type
of system for SN~Iax before the SN~2012Z progenitor discovery.
\citet{Liu10} present a model (though intended to
explain a different kind of system) that starts with a 7 \msun\ + 4
\msun\ close binary that undergoes two phases of mass transfer and
common envelope evolution and results in a 1 \msun\ C/O WD + 2 \msun\ He
star. As the He star evolves, it can again fill its Roche lobe and
begin stable mass transfer onto the white dwarf (at a high accretion
rate,  $\sim 10^{-5}$ \msun\ yr$^{-1}$, for instance) that could lead
to the SN~Iax. 

\index{Chandrasekhar limit}
\index{Chandrasekhar mass}
The low ejecta velocities of SN~Iax imply lower kinetic energy compared to
SN~Ia (under the reasonable assumption that SN~Iax do not have significantly
higher ejecta mass). Their lower luminosity also points in this direction
(though that depends specifically on how much $^{56}$Ni is synthesized).
Moreover, there is a much larger \emph{variation} in the kinetic energy in
SN~Iax. All of this points towards a deflagration (subsonic) explosion; pure
deflagration models of $M_{\rm Ch}$ C/O WDs show convoluted structure from the
turbulent flame propagation \citep[e.g.,][]{Gamezo03,2007ApJ...668.1118T} that
can produce a wide range of explosion energies. This contrasts with $M_{\rm
Ch}$ detonation scenarios that lead to more uniform energy release and a
layered structure \citep{Gamezo04,Gamezo05}.

\index{deflagration}
\index{pure deflagration}
Deflagrations are thought to naturally occur in the onset of runaway carbon
burning for $M_{\rm Ch}$ C/O WD progenitors \cite[e.g.,][ and references
therein]{2011ApJ...740....8Z,2012ApJ...745...73N}. Indeed for years, a leading
model to match observations of normal SN~Ia has been the delayed detonation
scenario, in which an initial deflagration transitions to a detonation after
the WD has expanded to lower density \citep{Khokhlov91,Gamezo05}. In SN~Iax,
one posits that this transition does not occur. Such a model matches many of
the observations \citep{2007ApJ...668.1132R,2013ApJ...771...58M}: lower yet
varied energy release, well-mixed composition \citep[this inhibits the
secondary near-infrared maximum;][]{Kasen06}, unusual velocity structure
\citep[e.g., Ca interior of Fe, something not seen in normal
SN~Ia;][]{Foley13}, and the strength of [\ion{Ni}{2}] in late-time spectra
implying stable nickel, preferentially produced at high density \citep[near
$M_{\rm Ch}$;][]{2016IJMPD..2530024M}. Early on, \citet{Branch04} suggested a
pure deflagration model to match spectra of SN~2002cx. In these models, there
is a prediction of significant unburned material (C/O), which may be
confirmed: carbon features are nearly ubiquitous in SN~Iax, more so than SN~Ia
\citep{Foley13}, and there are hints of low-velocity oxygen in some late-time
spectra of SN~Iax \citep{Jha06}. There remain challenges to a
pure-deflagration model of SN~Iax; for example, \citet{2015ApJ...805..150F}
find generically too-weak deflagrations for $M_{\rm Ch}$ progenitors because
of non-central, buoyancy driven ignition. Furthermore, asymmetry in pure
deflagrations may lead to polarization signatures in excess of what is
observed for SN~2005hk \citep{Chornock06,Maund10,2017ApJ...841...62M}.

\index{bound remnant}
The low-velocity late-time features, which have final velocities that can be
much less than the typical escape velocity from the surface of a white dwarf,
suggest that perhaps not all of the material was unbound. Similarly, the range
of energies that deflagrations produce could include outcomes less than the
binding energy. Thus, in this model, though SN~Iax would be $M_{\rm Ch}$
explosions, the ejecta mass could be significantly less, and the explosions
could leave behind a bound remnant. The luminous remnant would be
super-Eddington, and could be expected to drive an optically thick wind. This
scenario might explain the high densities inferred at late times in SN~Iax
spectra and why they do not become completely nebular \citep[$n \sim 10^9$
cm$^{-3}$;][]{McCully14}. Different amounts of ejecta versus wind material
could furthermore explain the diversity in line widths and strengths between
permitted and forbidden lines at late times \citep{Foley16}. A wind
``photosphere'' may also explain the luminous, red source seen in
post-explosion observations of SN~2008ha \citep{Foley14} and perhaps is
involved in producing the infrared excess in SN~2014dt
\citep{2016ApJ...816L..13F}. \citet{2017ApJ...834..180S} present an intriguing
model for these radioactively powered winds and show good agreement with
SN~Iax observations.

\subsection{Specific models for SN~Iax}

In some sense, the pure-deflagration $M_{\rm Ch}$ model is a ``failed'' SN~Ia;
indeed the primary reason these models were first explored was to explain
normal SN~Ia. However, first in 1-d and later in 3-d, it was shown these were
unlikely to match observations of SN~Ia (see references listed above).
After the identification and rapid observational growth in the
SN~Iax class, it became clear that these ``failures'' might be successfully
applied to SN~Iax.

\begin{figure}
\includegraphics[width=\textwidth]{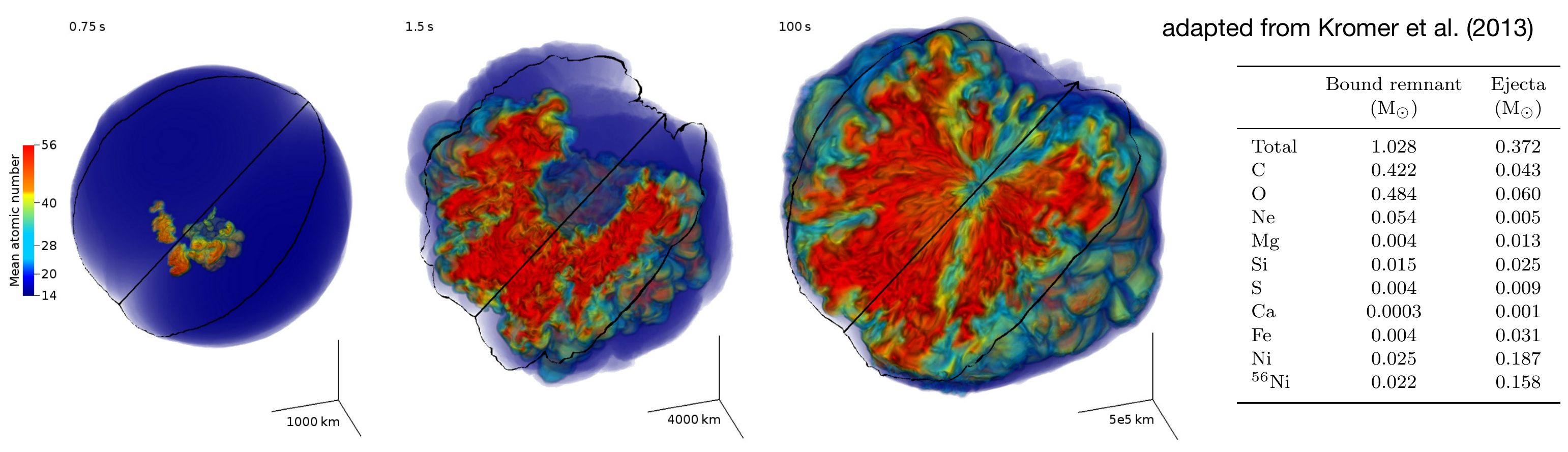}
\caption{Snapshots of a partial 3-d deflagration of a
Chandrasekhar-mass carbon-oxygen white dwarf that leaves a bound remnant. Note
the lack of burned material in the white dwarf core in the middle panel (1.5
sec into the explosion). While the thermonuclear runaway in the outer parts of
the white dwarf surrounds and engulfs the core at later times, the explosion
energy is not sufficient to unbind the star. The model predicts the
composition of the ejecta and remnant in the table shown. This figure is
adapted from \citet{Kromer13}.}
\label{fig:kromer}
\end{figure}

\citet{Jordan12} and \citet{Kromer13} presented 3-d simulations of
pure-deflagration $M_{\rm Ch}$ C/O WD explosions that did not fully unbind the
star (Figure \ref{fig:kromer}), and connected these to observed properties of
SN~Iax, as described above. \citet{Fink14} explore a range of initial
conditions in this scenario varying the number and location of ignition spots
and find that they can yield a wide range of total energy and ejecta masses
from 0.1 \msun\ to $M_{\rm Ch}$ (i.e., complete disruption). This provides a
natural way to explain the diversity observed in SN~Iax, and \citet{Magee16}
show some success in matching a particular realization of this explosion model
to observations of SN~2015H. \citet{2014ApJ...789..103L} also show
qualitatively similar results in being able to reproduce SN~Iax properties,
but find the opposite sense in the relation between energy yield and the
number of ignition spots, with high luminosity objects resulting from fewer
ignition points.

\index{hybrid white dwarfs}
A new wrinkle on the pure-deflagration $M_{\rm Ch}$ scenario is based the idea
of ``hybrid'' C/O/Ne white dwarfs
\citep{2013ApJ...772...37D,2014MNRAS.440.1274C}. Uncertainties in convective
mixing and carbon-flame quenching may allow for central carbon to exist in WDs
as massive as 1.3 \msun, rather than the traditional $\sim$1.05 \msun\
boundary between C/O and O/Ne white dwarfs
\citep{1984ApJ...277..791N,2017arXiv170306895D}. Such massive white dwarfs
come from more massive progenitors and require less accreted material to reach
$M_{\rm Ch}$: both of these lead to shorter delay time between formation and
explosion (perhaps as low as 30 Myr), and thus could be particularly relevant
to SN~Iax \citep{2014ApJ...789L..45M,2014ApJ...794L..28W}. Furthermore, a
range of masses for the C/O core could play a role in SN~Iax diversity
\citep{2015MNRAS.447.2696D,2015MNRAS.450.3045K}. \citet{2016A&A...589A..38B}
suggest that even delayed detonations of hybrid white dwarfs could explain
SN~Iax, though \citet{2016ApJ...832...13W} find the detonation phase makes
these explosions more similar to normal SN~Ia. \citet{2015ApJ...808..138L}
argue that from a binary evolution point of view the companion star to
SN~2012Z is best explained in a system with a C/O/Ne white dwarf primary.
\citet{2017arXiv170306895D} note some concerns about the viability of the
hybrid white dwarf scenario, including whether the carbon flame can be
successfully quenched \citep{2016ApJ...832...71L} or if the central C/O region
can survive without being mixed into the O/Ne layer above
\citep{2017ApJ...834L...9B}.

\index{double detonation}
Another class of SN~Iax models explores the recent resurgence in sub-$M_{\rm
Ch}$ double-detonation scenarios for normal SN~Ia. In this case varying WD
mass at explosion can lead to diversity, and perhaps explain both prompt SN~Ia
and the full range of SN~Iax
\citep{Wang13,2014RAA....14.1146Z,2016AA...589A..43N}. \citet{Stritzinger15}
argue that the ejecta mass for SN~2012Z is consistent with $M_{\rm Ch}$ and
advocate a pulsational delayed detonation model \citep{1995ApJ...444..831H}
for bright SN~Iax. \citet{2012MNRAS.419..827M} and \citet{2013ApJ...763..108F}
explore the potential of SN~Iax to result from white dwarf plus neutron star
mergers.

\section{Conclusion}
\label{sec:conc}

Taken together, observations and theory point to a leading model
that needs to be tested: \emph{a type Iax supernova results from a C/O or
(hybrid C/O/Ne) white dwarf that accretes helium from a He-star companion,
approaches the Chandrasekhar mass, and explodes as a deflagration that does
not necessarily completely disrupt the star.} It is quite possible, even
likely, that one or more aspects of this model is wrong, but it nonetheless
gives observers something to directly test and modelers a general framework to
explore and pick apart. Even within this model, there are important questions:
is $M_{\rm Ch}$ always required? Is it always a deflagration? Does varying the
ejecta/remnant mass explain the diversity? Does the WD + He-star channel
always lead to a SN~Iax?

\index{helium star}
In addition to testing this and other models with a broad range of
observations, we can look forward to some novel possibilities. For example,
what should happen to the He-star companion of SN~2012Z
\citep[e.g.,][]{2013ApJ...765..150S,2013ApJ...778..121L}? What might we expect
to observe from SN~Iax bound remnants and what happens to them? Will future
extremely-large-telescopes be able to spectroscopically confirm that the
companion to SN~2012Z ($m \approx 27.5$) was actually a helium star?

\index{helium nova}
What are the broader impacts of our understanding of SN~Iax? Is there a
connection to systems like the Galactic ``helium nova'' V445 Pup, a near
$M_{\rm Ch}$ white dwarf accreting from a helium star \citep{Kato08,Woudt09}?
Is it a SN~Iax precursor, or is some parameter different (e.g., the accretion
rate) that leads to the nova outcome?  Is there a connection between SN~Iax
and 2002es-like SN \citep{Ganeshalingam12}? Those are found in older
environments, but also have a detection of a likely single-degenerate
progenitor \citep{Cao15}. What is the relation of SN~Iax the population of
Ca-rich transients \citep{Perets10}? Are those C/O + He WD mergers?

Of course, one of the key reasons why understanding SN~Iax is important is the
insight that gives us about normal SN~Ia. But what is that insight telling us?
Does it point to sub-$M_{\rm Ch}$ or double degenerate progenitors for SN~Ia?
Is some form of detonation required in SN~Ia? Does the single degenerate
channel always lead to peculiar supernovae? One factor in explaining why the
SN~Ia progenitor/explosion problem has been with us for decades is the vast
array of possibilities to explode white dwarfs. Given the enormous effort that
has gone into explaining normal SN~Ia, it is astounding to think that we may
have a better understanding of their peculiar cousins, SN~Iax.

\section*{Cross-References}

Combustion in Thermonuclear Supernova Explosions \citep{Ropke17}\\
Evolution of Accreting White Dwarfs to the Thermonuclear Runaway \citep{Starrfield2016}\\
Explosion Physics of Thermonuclear Supernovae and Their Signatures
\citep{Hoeflich2017} \\
Light Curves of Type I Supernovae \citep{Bersten2017} \\
Low- and Intermediate-Mass Stars \citep{Karakas2017} \\
Nucleosynthesis in Thermonuclear Supernovae \citep{Seitenzahl2017} \\
Observational and Physical Classification of Supernovae \citep{Gal-Yam2017} \\
Population Synthesis of Massive Close Binary Evolution \citep{Eldridge2017} \\
Spectra of Supernovae During the Photospheric Phase \citep{Sim2017} \\
Supernova Progenitors Observed with HST \citep{VanDyk2016} \\
The Extremes of Thermonuclear Supernovae \citep{Taubenberger17} \\
Type Ia Supernovae \citep{Maguire2016} \\
Unusual Supernovae and Alternative Power Sources \citep{Kasen2017} \\

\begin{acknowledgement} 

I thank Ryan Foley and Curtis McCully for our close collaboration working on
SN~Iax and their assistance with this manuscript. I am also grateful to Mark
Sullivan, Daniela Graf, and Kerstin Beckert, for their seemingly inexhaustible
patience. This work was supported in part by US National Science Foundation
award 161545 and benefited greatly from discussions at the Munich Institute
for Astro- and Particle Physics Scientific Program ``The Physics of
Supernovae'' and associated Topical Workshop ``Supernovae: The Outliers.''
I dedicate this review to the memory of my friend and colleague, Weidong Li,
who started it all.

\end{acknowledgement}

\bibliography{handbookIax}

\end{document}